\documentclass{nature}

\usepackage{lineno}
\usepackage{amsmath}
\usepackage{amssymb}
\usepackage{graphicx}

\graphicspath{{./}{figures/}}

\title{Smoking-gun evidence for hierarchical black-hole mergers}

\begin{document}

\maketitle
\author{Yin-Jie~Li$^{1}$, Yuan-Zhu~Wang$^{2}$, Shao-Peng~Tang$^{1}$, Yi-Zhong~Fan$^{1,3,\ast}$.}

\begin{affiliations}
\small
\item{Key Laboratory of Dark Matter and Space Astronomy, Purple Mountain Observatory, Chinese Academy of Sciences, Nanjing 210023, People's Republic of China}
\item{Institute for Theoretical Physics and Cosmology, Zhejiang University of Technology, Hangzhou 310032, People's Republic of China}
\item{School of Astronomy and Space Science, University of Science and Technology of China, Hefei, Anhui 230026, People's Republic of China}\\
$^\ast$Corresponding author. Email: yzfan@pmo.ac.cn
\end{affiliations}

\begin{abstract}
How stellar-mass black holes grow after their birth is a central open question in astrophysics.
Gravitational-wave observations have revealed a subpopulation of coalescing black holes with both high masses and high spins, but whether these properties arise from hierarchical mergers in dense stellar environments or from accretion onto isolated black holes has remained unresolved.
Here, using a flexible mixture population model applied to the 259 binary black hole mergers in GWTC-5, we show that the mass function of the high-spin subpopulation traces, peak by peak, the predicted remnant-mass distribution of the low-spin, stellar-collapse-origin subpopulation up to $\sim80\,M_\odot$.
This morphological match, quantified by a Bhattacharyya coefficient as high as $\sim0.95$, is naturally expected if the high-spin black holes are themselves the products of earlier mergers, whereas any alternative scenario would require fine-tuning, thereby providing smoking-gun evidence for hierarchical mergers.
In addition, the sharp upper-mass cutoff of the low-spin subpopulation at $m_{\rm max,1}=54.2^{+7.7}_{-7.2}\,M_\odot$ yields an astrophysical $S$-factor of $S_{300}=151^{+30}_{-26}$~keV~b (68\% credible interval) for the $^{12}{\rm C}(\alpha,\gamma)^{16}{\rm O}$ reaction, in agreement with the benchmark theoretical value.
These results establish that the entire observed black-hole population can be accounted for by stellar collapse followed by dynamical hierarchical assembly, without invoking primordial black holes.
\end{abstract}

%%%%%%%%%%%%%%%%%%%%%%%%%%%%%%%%%%%%%%%%%%%%%%%%%%%%%%%%%%%%%%%%%%%

The successful operation of the LIGO/Virgo/KAGRA network has enabled the detection of hundreds of coalescing black holes, mostly in binary black hole (BBH) systems \cite{2016PhRvL.116f1102A, 2024PhRvD.109b2001A, 2023PhRvX..13d1039A, 2025arXiv250818082T, 2026arXiv260527225T}. 
A central question is how these binary black hole systems were formed and then evolved. 
In the literature, multiple formation channels are predicted (see ref.~\cite{2022PhR...955....1M} and the references therein).
The mixing of these channels smears out the signatures of individual subpopulations, complicating efforts to disentangle their respective formation and evolutionary histories.
For instance, hierarchical mergers arising from dynamical channels in dense environments are expected to populate the (pulsational) pair-instability supernova (PISN) mass gap \cite{2021NatAs...5..749G}, making it challenging to unambiguously identify such a feature and hence probe the $^{12}{\rm C}(\alpha,\gamma)^{16}{\rm O}$ reaction rate \cite{2020ApJ...902L..36F, 2021ApJ...912L..31W, 2022ApJ...924...39M, 2026RAA....26g5011X}.
A reliable determination of this gap requires robustly separating the first-generation (i.e., the stellar origin) and (potential) hierarchical subpopulations.

Current gravitational wave signals just tell us about BBHs several seconds (or even sub-second) before and after the mergers, which makes it challenging to reveal their evolution paths, particularly the hierarchical mergers, unless we hear them merge twice (e.g., the hierarchical triple mergers \cite{2024ApJ...965...80G}).
Moreover, the delay times are $\sim{\rm Myr}$--${\rm Gyr}$ long for individual events.
This issue can therefore only be addressed through population analysis, if we can reveal that a subpopulation comprises the remnants of previous mergers.
Higher-generation BHs as the remnants of previous mergers have a typical dimensionless spin magnitude of $\chi \sim 0.7$ \cite{2002MNRAS.330..232C, 2015MNRAS.454.3150G, 2021NatAs...5..749G, 2022PhR...955....1M}, and thus encode typical imprints (via effective/precession spin) in GW signals when they merge with other BHs \cite{2021PhRvD.104h4002B, 2024ApJ...966L..16P}.
The joint mass-spin distribution modeling of the coalescing black holes has been widely adopted to resolve the stellar-collapse and hierarchical merger subpopulations.
With the GWTC-3 and GWTC-4 data, it has been established that the coalescing black holes can be divided into two categories \cite{2022ApJ...941L..39W, 2024PhRvL.133e1401L, 2024arXiv240601679P, 2025PhRvL.134a1401A, 2025ApJ...987...65L, 2026Natur.652..874T, 2026NatAs.tmp..111A, 2025arXiv250915646B, 2025arXiv250923897L, 2026SCPMA..6999562W, 2026arXiv260107908P}.
One category is characterized by the low spin as well as a cutoff in the mass spectrum that is broadly consistent with the lower edge of PISN mass gap.
The other category covers a wider mass range and is distinguished by the high spin magnitudes. 
These features are generally consistent with the properties of the black holes from stellar-collapse and hierarchical merger, respectively \cite{2022ApJ...941L..39W, 2024PhRvL.133e1401L, 2024arXiv240601679P, 2025ApJ...981..177L, 2025PhRvL.134a1401A, 2025ApJ...987...65L, 2026Natur.652..874T, 2026NatAs.tmp..111A, 2025arXiv250915646B, 2025arXiv250923897L, 2026SCPMA..6999562W, 2026arXiv260107908P}.
Analyses of individual events such as GW241011 \cite{2025ApJ...993L..21A} lend further support to the hierarchical merger origin identified in population-level studies.
Firmly establishing these two distinct origins, however, is far more challenging.
This is particularly true for the hierarchical-merger origin, since accretion can also efficiently enhance both the mass and the spin of a black hole.
Indeed, it has been argued in the literature that the high-spin subpopulation of the coalescing black holes can also be formed via stellar evolution and/or accretion \cite{2026arXiv260628293R, 2025ApJ...994L..37K, 2026A&A...709A...5R, 2026arXiv260509351B, 2021ApJ...921L...2O, 2022ApJ...930...26S}.
It has even been claimed that the accretion model can better account for some aspects of the current data, for instance, the rather high spin with $\chi>0.8$ \cite{2026arXiv260509351B}.
The {\it critical} distinction is as follows: hierarchical mergers produce remnants whose masses are deterministically linked to the progenitor mass function, so that peaks in the first-generation spectrum map onto corresponding peaks in the second-generation spectrum.
By contrast, accretion-driven mass and spin growth depends on the local gas environment rather than on the progenitor mass distribution, so it generically does {\it not} reproduce the spectral morphology of the progenitor remnants.
This difference provides an {\it unambiguous, model-independent diagnostic}, which is the focus of this work.

In practice, the observed BBH mass function already presents prominent low-mass peaks \cite{2026arXiv260527226T}. Since hierarchical mergers directly reuse remnants from prior coalescences, these over-densities are naturally inherited by the hierarchical subpopulation: the hierarchical-merger mass function is expected to closely trace the final-mass distribution of remnants from the first-generation stellar-origin BBH subpopulation. Comparing these two distributions thus provides a powerful, nearly model-independent probe of successive BH merger chains. An additional practical advantage is that BBH masses are measured far more precisely than spins, making mass-domain features robust tracers of formation channels.
The newly released GWTC-5 catalog has substantially expanded the detected BBH sample \cite{2026arXiv260527225T}.
In this work we take a very flexible mixture population model to analyze these data. Our most intriguing finding is the {\it striking} similarity between the mass distribution of the high-spin subpopulation and that of remnant BHs from mergers in the low-spin stellar-origin subpopulation, a result that not only provides direct evidence for the high-spin subpopulation's hierarchical merger origin, but also sheds valuable light on the dynamical formation channels of the low-spin subpopulation.

\begin{figure}
\centering  
\includegraphics[width=0.99\linewidth]{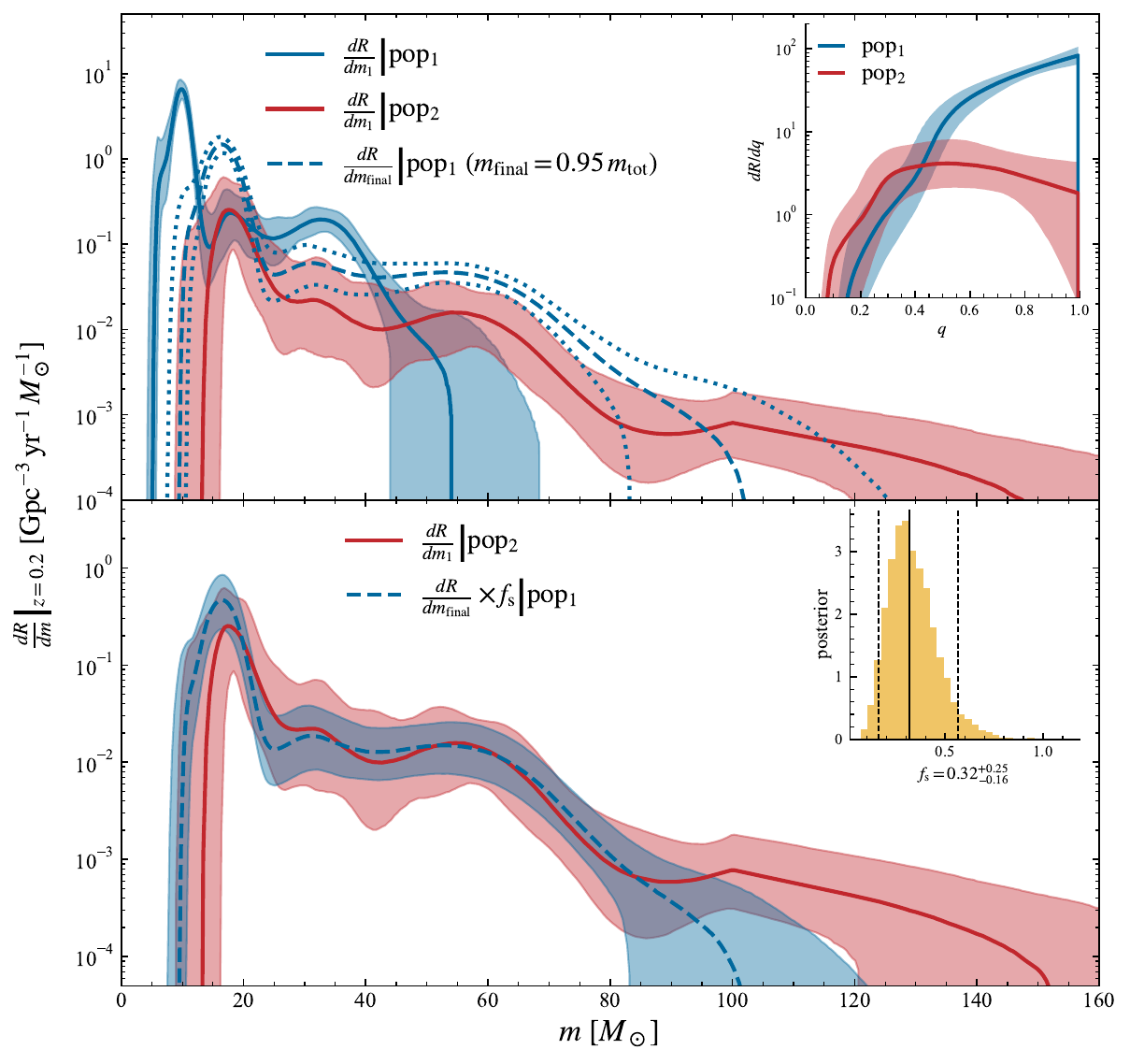}
\caption{The reconstructed mass distributions of the black hole groups. The solid curves are for the medians and the regions represent the 90\% credible intervals. 
Top: the low and high spin subpopulations are marked in blue and red, and the dashed lines represent the reconstructed mass distribution of remnant black holes formed in the mergers of the low-spin subpopulation objects (i.e., the remnant group, whose mass has been approximated as $0.95(m_1+m_2)$).
Bottom: the comparison of the reconstructed mass distribution of the second subpopulation (in red) with that of the remnant population but scaled by a factor of $\sim 0.32^{+0.25}_{-0.16}$ (90\%C.L.).}
\label{fig:m1_mf}
\end{figure}

\textbf{\emph{Results}}

We employ the standard hierarchical Bayesian inference framework for GW population studies \cite{2019MNRAS.486.1086M, 2024PhRvD.109b2001A, 2024PhRvL.133e1401L, 2026arXiv260527226T} to extract these features from GWTC-5. Each event delivers noisy posteriors for parameters $(m_1,m_2,\chi_{\rm eff},\chi_{\rm p},z)$. This approach combines the full catalog to recover intrinsic astrophysical distributions, rigorously accounting for measurement uncertainties and selection effects (preferential detection of more massive, closer binaries). We define a population model $\pi(\lambda|\mathbf{\Lambda})$ governed by hyper-parameters $\mathbf{\Lambda}$, and update their posteriors via the full catalog likelihood after marginalizing over individual-event uncertainties (see the Methods for details). Hereafter, we report 90\% credible intervals unless otherwise stated.

We apply a flexible mixture population model describing $\chi_{\rm eff}-\chi_{\rm p}-m_1-q-z$ distribution of BBHs, mainly adopted from our earlier work \cite{2026SCPMA..6999562W}, to perform hierarchical Bayesian inference on the GWTC-5 dataset.
Notably, the present approach achieves greater flexibility, particularly in the mass functions, by employing two \textsc{PowerLawSpline} functions\cite{2022ApJ...924..101E}, each paired with a distinct mass-ratio distribution.
Full details of the inference method and the population model are provided in the Methods.
We confidently resolve two subpopulations with clear structures in their mass functions, see the top panel of Figure~\ref{fig:m1_mf}, and rule out a single population with a Bayes factor of $\ln\mathcal{B}=41$.

Figure~\ref{fig:m1_mf} shows the inferred mass functions of the subpopulations. 
In the top panel, the blue and red regions are for low and high spin subpopulations, and the dashed/dotted curves are for the reconstructed mass distribution of remnant black holes formed in the mergers of the low-spin subpopulation objects (i.e., the remnant population).
Features/structures emerge in the mass functions for both the low-spin subpopulation (pop$_1$) and the high-spin subpopulation (pop$_2$), see also Extended Figure~\ref{fig:spline}.
These features enable us to further reveal the relationship between the two subpopulations.
We therefore generate the final-mass (which is approximated as $m_{\rm final}=0.95m_{\rm tot}$, where $m_{\rm tot}=(m_1+m_2)$, although unequal systems may radiate $\lesssim 5\%$ total mass during mergers) distribution of the pop$_1$ (dashed curve/band) and compare it to the primary mass distribution of the pop$_2$.
To better view the consistency/difference of the shapes of these two mass functions, we define a scale factor of 
$f_{\rm s}\equiv{\int_{40}^{80}dm\frac{{d} \mathcal{R}}{dm_{1}}|_{\rm pop_2}}/{\int_{40}^{80}dm\frac{{d} \mathcal{R}}{dm_{\rm final}}|_{\rm pop_1}} = 0.32^{+0.25}_{-0.16}$ in $[40,80]M_\odot$
(this range is chosen because most high-spin events concentrate here, yielding more precise rate measurements; the results are robust against alternative choices of the lower boundary, see the Supplementary Information for details).
We therefore scale the final-mass function of the low spin population (pop$_{1}$) with $f_{\rm s}$, see bottom panel of Figure~\ref{fig:m1_mf} (blue dashed curve).
It is well consistent with the mass function of the pop$_2$ (red curve) below $\sim 80M_\odot$. 

\begin{figure}
\centering  
\includegraphics[width=0.8\linewidth]{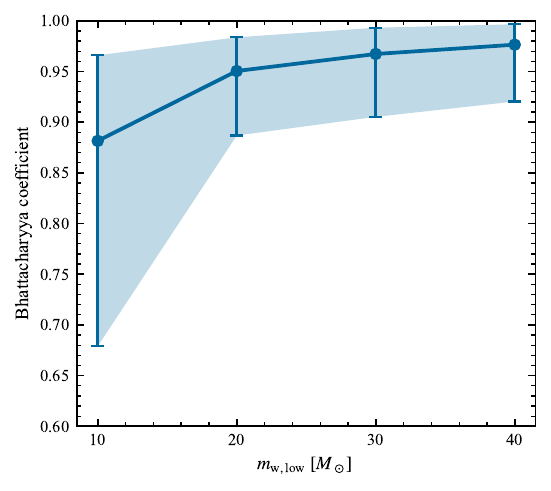}
\caption{The Bhattacharyya coefficient as a function of the lower edge of the mass window ($m_{\rm w,low}$, $80~M_\odot$).
This coefficient \cite{Bhattacharyya1943} describes the shape similarity among ${\frac{d\mathcal{R}}{dm_{1}}|_{\rm pop_2}}$ and ${\frac{d\mathcal{R}}{dm_{\rm final}}|_{\rm pop_1}}$.
The shaded regions represent the medians and $90\%$ credible intervals.}
\label{fig:BC_test}
\end{figure}

\paragraph{Direct evidence for hierarchical-merger origin of the high-spin subpopulation.}
The almost identical shapes of ${\frac{{d} \mathcal{R}}{dm_{1}}|_{\rm pop_2}}$ and ${\frac{{d} \mathcal{R}}{dm_{\rm final}}|_{\rm pop_1}}$ in a wide mass range, as demonstrated in the bottom panel of Figure~\ref{fig:m1_mf}, is the main finding of this work.
We further quantify this similarity by comparing the two mass functions in shape only, i.e., after normalizing each to unit integral within a common mass window, so the uncertainty introduced by $f_{\rm s}$ is avoided.
As shown in Figure \ref{fig:BC_test}, in a wide range of $m_{\rm w,low}$ (the lower edge of the mass window), the Bhattacharyya coefficient \cite{Bhattacharyya1943} for ${\frac{d\mathcal{R}}{dm_{1}}|_{\rm pop_2}}$ and ${\frac{d\mathcal{R}}{dm_{\rm final}}|_{\rm pop_1}}$ is close to $1$, confirming that the two mass distributions are statistically well consistent in the shape.
Same conclusion is drawn for other tests (see the Supplementary Information).

Such a remarkable similarity sheds valuable light on the intrinsic correlation of these two subpopulations.
The straightforward interpretation is that the high-spin massive subpopulation (pop$_{2}$) has a hierarchical-merger origin, i.e., these objects were generated by the mergers of stellar collapse-origin black holes (pop$_{1}$).
This requires either that stellar-collapse-origin black holes, regardless of their birth sites, share a universal mass distribution, or that most merger products of pop$_{1}$ (particularly those with $m>30\,M_\odot$) are born in dense environments and undergo further mergers (unless ejected).
Any alternative interpretation would need to independently reproduce the same multi-peaked spectral morphology, including the relative heights and positions of at least two peaks without a causal link to the first-generation mass function, which constitutes a fine-tuning of at least several free parameters.
We therefore conclude that the second subpopulation is dominated by the hierarchical mergers of the pop$_{\rm 1}$ black holes. We obtain an overall merger fraction (or retention fraction) of the remnant as $f_{\rm merger}\equiv \frac{r_2}{(1-r_2)/2}=23^{+17}_{-11}\%$ and $f(m)_{\rm merger}\sim30\%$ at the mass $\gtrsim30M_\odot$ (see Extended Data Figure~\ref{fig:ratio_function}), where $r_{\rm 2}$ is the mixture fraction of pop$_2$.
Considering the merger rate evolves as $\sim(1+z)^{2.5}$ (see Ref.\cite{2026arXiv260527226T} also Supplementary Figure~\ref{fig:gamma}), the inclusion of the delay time between the first-generation and hierarchical mergers would lower the inferred $f_{\rm merger}$.
For short ($t_{\rm del}\sim0.1$--$0.5$~Gyr), medium ($\sim1$--$3$~Gyr), and long ($\sim5$--$10$~Gyr) delay channels \cite{2018ApJ...866L...5R}, $f_{\rm merger}$ would decrease by $\lesssim10\%$, $\sim20$--$50\%$, and $\sim60$--$90\%$, respectively.

\paragraph{Tight constraint on the `Holy Grail' reaction.}
The cutoff of the mass function of the stellar-collapse-origin black holes (i.e., $m_{\rm max,1}$) is a key parameter, since it provides a valuable probe of the cross section of the so-called `Holy Grail' reaction (i.e., $^{12}{\rm C}(\alpha,\gamma)^{16}{\rm O}$).
As already noted in ref.~\cite{2026arXiv260527226T}, the O4b data have a much softer mass function at $m_1>50M_\odot$ than the previous data.
As a result, the mass cutoff of pop$_{1}$ inferred with the current GWTC-5 data shifts to $m_{\rm max,1}=54^{+14}_{-10}\,M_\odot$ (90\%\ credibility), which is lower, though still comparable with the value of $63^{+23}_{-20}M_\odot$ found with the GWTC-4 data for a similar mass distribution model \cite{2026SCPMA..6999562W}.
Benefiting from the significantly reduced uncertainty, we now infer $S_{300}=151_{-26}^{+30}\ \text{keV b}$ ($68\%$ credibility), assuming that $m_{\rm max,1}$ represents the lower edge of the PISN mass gap.
This is well consistent with the theoretically recommended value of $140\pm21$ keV b (68\% credibility) \cite{2017RvMP...89c5007D}.
In the next decade, $m_{\rm max,1}$ will be more precisely measured, with which the accuracy of the inferred $S_{300}$ will be comparable with or better than that of the theoretical inference.

\begin{figure}
\centering
\includegraphics[width=0.99\linewidth]{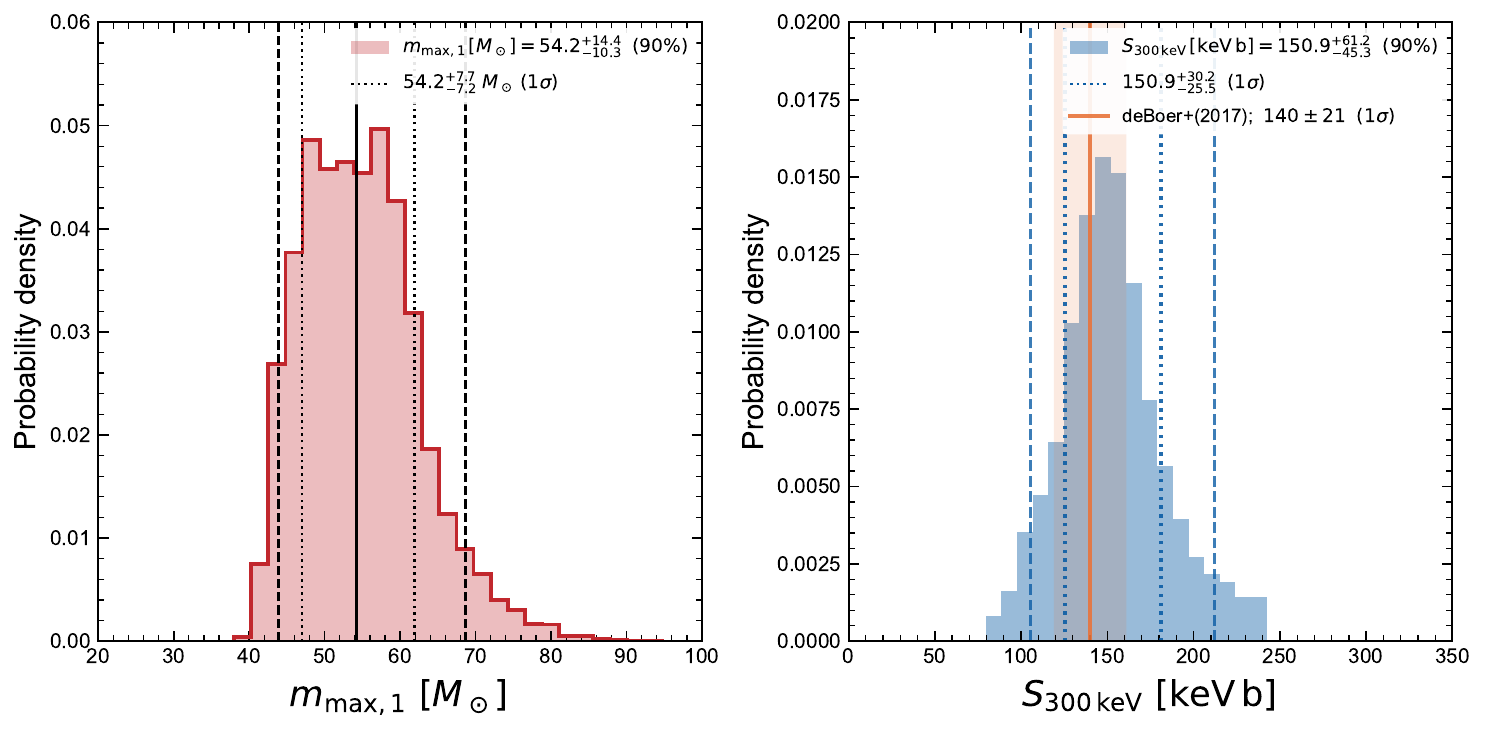}
\caption{{\bf The inferred $S_{\rm 300keV}$.} The evaluated $S$ factor of $^{12}{\rm C}(\alpha,\gamma)^{16}{\rm O}$ reaction at 300 keV assuming that $m_{\rm max,1}$ represents the lower edge of the PISN mass gap. The value of $S_{\rm 300keV}$ recommended by ref.~\cite{2017RvMP...89c5007D} (1$\sigma$ region including the model uncertainty) is also plotted for comparison.}
\label{fig:S300}
\end{figure}

\paragraph{Features of the second subpopulation.}
In the mass function of pop$_{1}$, we recover two peaks, at $\sim10\,M_\odot$ and $\sim35\,M_\odot$, and a dip (possibly a gap) near $\sim14\,M_\odot$, consistent with previous analyses (e.g., Refs. \cite{2021ApJ...913L..19T, 2022ApJ...924..101E, 2026arXiv260401420L}). 
More importantly, the mass function of pop$_{2}$ shows two (and possibly three) peaks at $\sim17\,M_\odot$ and $\sim 58\,M_\odot$.
The over-density at $m_1\simeq58\,M_\odot$ is established at the $>99\%$ credible level (see Extended Data Figure~\ref{fig:spline}).
Although the $\sim17M_\odot$ peak is not significant in the spline function (which is reduced by the turn-on smooth function), the peak in the mass function is evident, similar to that of the first peak in the mass function of pop$_{1}$.
We find the presence of a high-mass peak for pop$_{2}$ is favored by a Bayes factor of $\ln\mathcal{B}=3$, with model comparison using the parametric models described in the Methods (Extended Data Table~\ref{tab:BF}).

The locations of these peaks match nicely with those of the final masses of $\text{pop}_{1}$. 
These features are highly informative about the evolution path of binary black holes.
The astrophysical origin of the $\sim 35\,M_\odot$ feature in the BH mass spectrum remains an important open question, as it may encode key information about massive-star evolution and the dominant formation channels. 
While originally interpreted as a signature of pair-instability physics \cite{2019ApJ...882...36M}, reproducing this feature at such relatively low masses is theoretically challenging \cite{2021ApJ...913...42W}, leading to the proposal of various alternative channels \cite{2026arXiv260407456G, 2025arXiv250915646B}.
We address this ambiguity by comparing the remnant distribution of first-generation mergers with that of the hierarchical-merger subpopulation, providing the first direct evidence that BBHs in the $\sim 35\,M_\odot$ over-density contribute significantly to the hierarchical-merger population, indicating that these mergers occur in dense environments such as star clusters and/or AGN disks.

\textbf{\emph{Discussion}}

To clarify the robustness of our main finding, i.e., the remarkable consistency of the shapes of ${d\mathcal{R}\over dm_{1}}|_{\rm pop_2}$ and $f_{\rm s}{d\mathcal{R}\over dm_{\rm final}}|_{\rm pop_1}$ (see the bottom panel of Figure ~\ref{fig:m1_mf}), we have carried out further tests.
First, we check our results with different thresholds of variance for likelihood estimation.
It turns out that the mass functions of the two subpopulations remain unchanged, when we adopt a more lenient threshold setting, although the $\chi_{\rm p}$ distribution becomes narrower, see Supplementary Figure~\ref{fig:varth}.
Second, we re-analyze the mass distributions of the subpopulations in two additional ways.
One is comparing the hierarchical-merger mass function to the $m_{\rm final}$ distribution directly inferred from data (rather than converting from the mass function of the first-subpopulation), see Supplementary Figure~\ref{fig:mfinal_direct}.
The other is an analysis with another population model for $m_1-m_2-\chi_1-\chi_2-\cos\theta_1-\cos\theta_2$ with a pairing function, as introduced in refs.~\cite{2025arXiv250923897L, 2026SCPMA..6999562W}, see Supplementary Figure~\ref{fig:mact}.
Both analyses show that $\left.\frac{d\mathcal{R}}{dm_1}\right|_{\rm pop_2}$ well traces $f_{\rm s}\left.\frac{d\mathcal{R}}{dm_{\rm final}}\right|_{\rm pop_1}$.
These tests demonstrate that our main finding is solid and it provides the direct evidence for the hierarchical merger origin of (most) high-spin coalescing black holes. 
%With direct evidence that the second subpopulation are hierarchical mergers of remnants of the first subpopulation, as illustrated above, we report that
The hierarchical mergers take a fraction of $r_2 = 10^{+6}_{-4}\%$, corresponding to a merger rate of $2.7^{+1.6}_{-1.2}~{\rm Gpc^{-3} yr^{-1}}$ at $z=0.2$.
This $r_2$ is higher than that reported in previous catalogs \cite{2024PhRvL.133e1401L, 2025arXiv250923897L}, as a result of the significant enrichment of hierarchical mergers in the low-mass regime in GWTC-5.

While the shapes of $\left.\frac{d\mathcal{R}}{dm_1}\right|_{\rm pop_2}$ and $f_{\rm s}\left.\frac{d\mathcal{R}}{dm_{\rm final}}\right|_{\rm pop_1}$ show broad consistency across a wide mass range, a new component, likely corresponding to higher-generation BHs, is required to account for deviations above $80M_\odot$ and possibly near $25M_\odot$. As shown in Extended Data Figure~\ref{fig:pop2}, the resulting $f_{\rm s}\left.\frac{d\mathcal{R}}{dm_{\rm final}}\right|_{\rm pop_2}$ of merging compact objects features a prominent peak at $\sim 25M_\odot$ and a high-mass tail extending beyond $80M_\odot$.
The existence of these higher-generation remnants implies that BHs in dense environments with deep gravitational potential wells can sustain multi-step mass growth via repeated mergers \cite{2026ApJ...999..127L,2025PhRvD.112f3034X}, a process that may seed intermediate-mass black hole formation. This provides long-sought observational evidence for the theoretically predicted hierarchical assembly pathway. In the precision era of merging BH mass function measurements, our approach opens a new window for probing BBH formation channels, particularly dynamical pathways in dense astrophysical environments.

Owing to the enlarged BBH sample, the subpopulation modeling yields an improved bound of $m_{\rm max,1}=54.2^{+7.7}_{-7.2}\,M_\odot$ (68\% credibility) for the low-spin BHs. 
Interpreting $m_{\rm max,1}$ as the lower edge of the PISN mass gap, we obtain $S_{\rm 300keV}=151^{+30}_{-26}$ keV b (68\% credibility) for the key reaction $^{12}{\rm C}(\alpha,\gamma)^{16}{\rm O}$, consistent with the benchmark theoretical value recommended in the literature \cite{2017RvMP...89c5007D}.
Thus, the low-spin subpopulation is fully consistent with a stellar-collapse origin.
Together with the direct evidence that the high-spin coalescing BHs originate from hierarchical mergers, there is no need for primordial BHs in interpreting the GWTC-5 data.
The tight bound on $m_{\rm max,1}$ has another important implication.
Slowly rotating black holes as massive as $\sim 93M_\odot$ can be generated in the single star evolution scenario \cite{2024MNRAS.529.2980W}.
Now the 95\% upper limit is $\approx 68M_\odot$, suggesting that a massive low-spin black hole subpopulation generated by single-star evolution, if it exists, has yet to be detected.

%*************************************

\newpage
\setcounter{figure}{0}
\setcounter{table}{0}
\renewcommand{\figurename}{Extended Data Figure}
\renewcommand{\tablename}{Extended Data Table}
\begin{center}
{\bf Methods}
\end{center}

\textbf{\emph{Hierarchical Bayesian inference}}

We perform hierarchical Bayesian inference using the GWTC-5 catalog \cite{2026arXiv260527225T}, which contains 259 BBH events with false alarm rate $<1$ yr$^{-1}$.
Posterior samples for each event are taken from the public release, and we use the same posterior samples as ref.~\cite{2026arXiv260527226T}.
The likelihood is
\begin{equation}
\mathcal{L}(\vec{d}\mid \mathbf{\Lambda}) \propto N^{N_{\rm obs}}\exp(-N\eta(\mathbf{\Lambda}))\prod_{i}^{N_{\rm obs}}\frac{1}{n_i}\sum_{k}^{n_i}\frac{\pi(\lambda_{i}^k\mid \mathbf{\Lambda})}{\pi(\lambda_{i}^k\mid \varnothing)},
\end{equation}
where $\eta(\mathbf{\Lambda})$ is the detection fraction, which is estimated using injection campaigns \cite{2025PhRvD.112j2001E}. 
We use the \textsc{Bilby} package with \textsc{PyMultinest} sampler.
Following ref.~\cite{2026arXiv260527226T}, we require the variance of likelihood estimator to be $<{\rm Var}_{\rm thr}=1$ for the results reported in this work; while a more relaxed threshold ($<{\rm Var}_{\rm thr}=4$) is also used to check the robustness of our results, see Supplementary Figure~\ref{fig:varth}.

\textbf{\emph{The fiducial population model}}

We adopt a flexible mixture model describing $m_1-m_2-\chi_{\rm eff}-\chi_{\rm p}$ distribution as our primary analysis, which allows (two) subpopulations to have different 4-D parameter distributions.
A 5-D mixture model (including different rate evolution model) is also tested, and the evidence that the high-spin/high-mass subpopulation evolves faster (so called `the heavier the faster' \cite{2024ApJ...975...54G}) is weakened, see Supplementary Figure~\ref{fig:gamma}.
We adopt another flexible mixture model for $m_1-m_2-\chi_1-\chi_2-\cos\theta_1-\cos\theta_2$ distribution with a pairing function of $\propto (m_2/m_1)^\beta$ as a cross-check model, such kind of models were previously used for GWTC-3 and GWTC-4 \cite{2024PhRvL.133e1401L, 2025arXiv250923897L, 2026SCPMA..6999562W}.
For both primary and cross-check models, the mass functions are flexible enough to catch the underlying mass distributions of subpopulations.

One of the reasons why we use $\chi_{\rm eff}-\chi_{\rm p}$ is that $\chi_{\rm eff}-\chi_{\rm p}$ space is two-dimensional, lower than the $\chi_1-\chi_2-\cos\theta_1-\cos\theta_2$ space for BBHs, although some information (about individual BHs) may be neglected.
It allows the model to carry structures in the mass function ensuring a reliable likelihood estimation (subject to Monte Carlo integration with limited posterior samples and injections). 
As shown in the Supplementary Materials of ref.~\cite{2024PhRvL.133e1401L}, more samples and loose threshold allow more significant features in distribution (see also Figures 15--17 of ref.~\cite{2026arXiv260527226T}).
Another reason is that the $m_1-m_2-\chi_{\rm eff}$-$\chi_{\rm p}$ allows different mass-ratio distributions for (two) subpopulations, however such flexibility is difficult to realize in the pairing function. 

We model the source-frame population distribution of $(m_1,m_2,\chi_{\rm eff},\chi_{\rm p},z)$ with a two-component mixture, building upon our previous works \cite{2025ApJ...987...65L, 2026SCPMA..6999562W}
\begin{equation}
\begin{split}
&p(m_1,m_2,\chi_{\rm eff},\chi_{\rm p},z\mid\Lambda)
= p(z\mid \gamma)\, \times \\
&\Big[(1-r_2)\,p_1(m_1)\,p_1(m_2\mid m_1)\,p_1(\chi_{\rm eff},\chi_{\rm p}) \\
&\;\, + r_2\,p_2(m_1)\,p_2(m_2\mid m_1)\,p_2(\chi_{\rm eff},\chi_{\rm p})\Big],
\end{split}
\end{equation}
where $r_2\in[0,1]$ is the mixture fraction of the second subpopulation and
\begin{equation}
p(z\mid\gamma)\propto (1+z)^{\gamma}.
\end{equation}
Note that the two subpopulations are assigned \emph{distinct} conditional mass-ratio distributions (see below), reflecting their different formation channels.
The priors of the fiducial model are summarized in Extended Data Table~\ref{tab:prior_xeff_polar_xp}.

\paragraph{Mass-ratio distributions.}
For the first (1g) subpopulation we adopt a power law,
\begin{equation}
p_1(m_2\mid m_1)\propto m_2^{\beta_1}\,,
\qquad m_2\in[m_{\min,1},m_1].
\end{equation}
For the second (hierarchical) subpopulation we instead use a truncated Gaussian
whose mean and width scale with the primary mass,
\begin{equation}
p_2(m_2\mid m_1)
=\mathcal{N}\!\left(m_2\mid \mu_q\,m_1,\ \sigma_q\,m_1\right)_{[m_{\min,1},\,m_1]},
\end{equation}
so that $\mu_q$ and $\sigma_q$ control, respectively, the typical mass ratio and
its dispersion in units of $m_1$. This allows the hierarchical population to
favor more symmetric (comparable-mass) pairings, as expected for repeated
mergers, while keeping the 1g population power-law-like.

\paragraph{Primary-mass distribution.}
For each subpopulation $i\in\{1,2\}$ we model the primary-mass distribution as
\begin{equation}
\begin{split}
p_i(m_1)\propto&~ m_1^{-\alpha_i}\,S(m_1\mid m_{\min,i},\delta_i)\,
\exp\!\big[f_i(m_1)\big],\\
&\quad m_1\in(m_{\min,i},m_{\max,i}),
\end{split}
\end{equation}
where $S(m_1\mid m_{\min,i},\delta_i)$ is a smooth turn-on at the lower edge, and $f_i(m_1)$ is a cubic-spline perturbation defined on a fixed knot grid.
We use
\begin{equation}\label{eq:smooth}
S(m\mid m_{\min},\delta)=
\left(1+\exp\!\left[\frac{\delta}{m'}+\frac{\delta}{m'-\delta}\right]\right)^{-1},
\end{equation}
with $m'\equiv m-m_{\min}$, for $m\in(m_{\min},m_{\min}+\delta]$, and $S=1$ for $m>m_{\min}+\delta$.
The spline $f_i(m_1)$ is specified by values $\{n_j\}$ (subpopulation 1) or $\{o_j\}$ (subpopulation 2) at logarithmically spaced knots, and interpolated with a natural cubic spline; the overall normalization of $p_i(m_1)$ is computed numerically.

\paragraph{Spin distribution of subpopulation 1.}
For the first (low-spin) subpopulation we assume
\begin{equation}
p_1(\chi_{\rm eff},\chi_{\rm p})
= p_1(\chi_{\rm eff})\,p_1(\chi_{\rm p}\mid \chi_{\rm eff}),
\end{equation}
with truncated Gaussians
\begin{equation}
p_1(\chi_{\rm eff})
= \mathcal{N}(\chi_{\rm eff}\mid \mu_{x,1},\sigma_{x,1})_{[-1,1]},
\quad \sigma_{x,1}=10^{\lg\sigma_{x,1}},
\end{equation}
and
\begin{equation}
\begin{split}
p_1(\chi_{\rm p}\mid \chi_{\rm eff})
&=\mathcal{N}\!\left(\chi_{\rm p}\mid \mu_{y,1}+\rho_1\chi_{\rm eff},\sigma_{y,1}\right)_{[0,1]},\\
\sigma_{y,1}&=10^{\lg\sigma_{y,1}}.
\end{split}
\end{equation}

\paragraph{Spin distribution of subpopulation 2 (polar semi-annulus).}
For the second subpopulation we introduce an elliptical polar mapping
\begin{equation}
x=\frac{\chi_{\rm eff}}{l_{\chi_{\rm eff}}},\quad
y=\frac{\chi_{\rm p}}{l_{\chi_{\rm p}}},\quad
r=\sqrt{x^2+y^2},\quad
t=\arctan2(x,y),
\end{equation}
and define, up to normalization on $\chi_{\rm eff}\in[-1,1]$ and
$\chi_{\rm p}\in[0,1]$,
\begin{equation}
p_2(\chi_{\rm eff},\chi_{\rm p})
\propto p_r(r)\,p_t(t),
\end{equation}
with a radial shell
\begin{equation}
p_r(r)=\exp\!\left[-\left(\frac{r-1}{\sigma_r}\right)^2\right],
\qquad \sigma_r = 10^{\lg\sigma_{\chi}},
\end{equation}
and an angular component
\begin{equation}
p_t(t)\propto \exp\!\big[f(t)\big]\,
{\bf 1}_{[t_{\min},t_{\max}]}(t)\,
\cos t.
\end{equation}
The function $f(t)$ is represented by a natural cubic spline with 6 knots uniformly spaced in $t\in[-\pi/2,\pi/2]$ and spline amplitudes $\{n_{x,j}\}_{j=1}^{6}$.
The normalization constant is computed numerically.
Note that $p_t(t)$ controls the skewness of the distribution, enabling us to identify structures beyond isotropic spins introduced in refs.~\cite{2021PhRvD.104h4002B, 2024ApJ...966L..16P}.
If $f(t)$ is fixed to a constant with $t_{\min}=-\pi/2$ and $t_{\max}=\pi/2$, then $\chi_{\rm eff}$ reduces to a uniform-like distribution, corresponding to typical star-cluster channels.
All the parameters and its priors are summarized in Extended Data Table~\ref{tab:prior_xeff_polar_xp}.

\paragraph{Parametric primary-mass model (P2P).}
As a cross-check against the flexible spline model, we adopt a fully parametric primary-mass distribution for each subpopulation $i\in\{1,2\}$, replacing the spline perturbation $\exp[f_i(m_1)]$ with a \textsc{PowerLaw2Peak} (P2P) form,
\begin{equation}
\begin{split}
p_i(m_1)\propto&~\Big[(1-r_{p,i})\,\mathcal{P}(m_1\mid\alpha_i,m_{\min,i},m_{\max,i}) \\
&+ r_{p,i}\big((1-r_{p2,i})\,\mathcal{N}_{1,i} \\
&\qquad\quad + r_{p2,i}\,\mathcal{N}_{2,i}\big)\Big]\,
S(m_1\mid m_{\min,i},\delta_i),
\end{split}
\end{equation}
where we abbreviate the truncated Gaussians as
\begin{align}
\mathcal{N}_{1,i}&\equiv\mathcal{N}(m_1\mid\mu_{1,i},\sigma_{1,i})_{[m_{\min,i},m_{\max,i}]},\\
\mathcal{N}_{2,i}&\equiv\mathcal{N}(m_1\mid\mu_{2,i},\sigma_{2,i})_{[m_{\min,i},m_{\max,i}]},
\end{align}
$\mathcal{P}(m_1\mid\alpha_i,m_{\min,i},m_{\max,i})\propto m_1^{-\alpha_i}$
is a normalized truncated power law on $(m_{\min,i},m_{\max,i})$, $S(m_1\mid m_{\min,i},\delta_i)$ is the same smooth turn-on as in Eq.~(\ref{eq:smooth}), $r_{p,i}\in[0,1]$ is the total fraction in the two Gaussian peaks, and $r_{p2,i}\in[0,1]$ partitions the peak weight between the two Gaussian components centered at $\mu_{1,i}$ and $\mu_{2,i}$.
The overall normalization of $p_i(m_1)$ is computed numerically.
The results obtained with the parametric primary-mass model are shown in Extended Data Figure~\ref{fig:P2P}.
We also use this parametric formula for model comparison, to validate the mass peaks in the second subpopulation, see Extended Data Table~\ref{tab:BF} for Bayes factors.

\textbf{\emph{Shape-consistency diagnostics}}

\begin{figure}
\centering  
\includegraphics[width=0.99\linewidth]{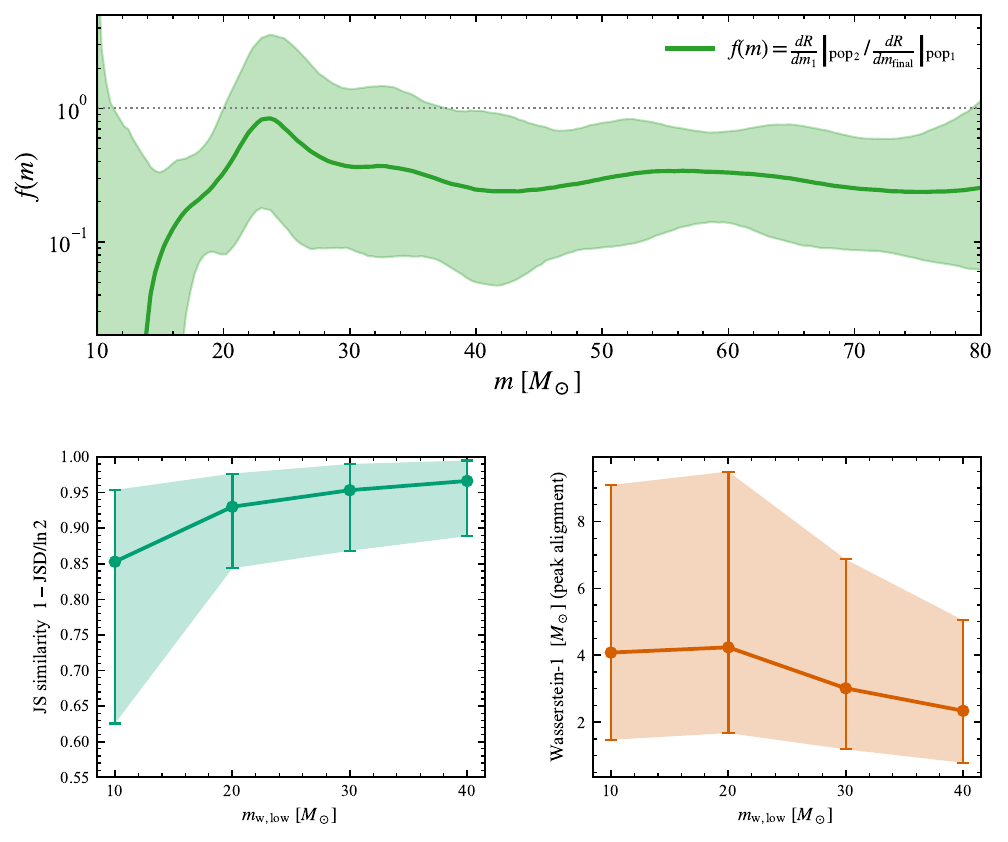}
\caption{Top: Ratio between pop$_2$ $m_1$ and pop$_1$ $m_{\rm final}$, as a function of mass, the solid lines and shaded bands are for median and 90\% credible intervals. Bottom: Shape similarity between the \textit{pop1} final-mass and \textit{pop2} primary-mass ($m_1$) distributions as a function of the lower window edge (upper edge fixed at $80~M_\odot$). Bottom panels show the Jensen--Shannon similarity (left), and the peak-aligned Wasserstein-1 distance (right); arrows indicate better agreement, markers denote posterior medians, and shaded bands show the $90\%$ credible intervals.}
\label{fig:ratio_function}
\end{figure}

We calculate the scale factor $f_{\rm s}$ within $[40,80]\,M_\odot$ because most high-spin events are located in this range (see Supplementary Figure~\ref{fig:informed}), so the measurement of the rate density is most precise there.
If the scale factor is instead computed over the full mass range, $f_{\rm s}$ reduces to $\frac{2r_2}{1-r_2}=24^{+17}_{-11}\%$, still consistent with $f_{\rm s,40\text{--}80}$.
An additional astrophysical consideration is that isolated field-evolution channels may contribute significantly in the low-mass range, and the origin of the possible
over-density of $\mathrm{pop}_2$ at $\sim30\,M_\odot$ remains uncertain (see the Discussion).
These considerations motivate our fiducial choice of $[40,80]\,M_\odot$.
As shown below, the shape consistency between the two spectra is nevertheless robust against this choice: adopting $[20,80]\,M_\odot$ or $[30,80]\,M_\odot$ yields compatible results (see also Extended Data Figure~\ref{fig:ratio_function} for the ratio as a function of mass).

To test whether the first-generation mass spectrum of $\mathrm{pop_2}$, $p(m)\!\equiv\!\left.\frac{d\mathcal{R}}{dm_1}\right|_{\mathrm{pop_2}}$, follows the hierarchical-merger product spectrum of $\mathrm{pop_1}$, $q(m)\!\equiv\!\left.\frac{d\mathcal{R}}{dm_{\rm final}}\right|_{\mathrm{pop_1}}$, we compare the two distributions in {\it shape} only: within each window $[m_{\rm w,low},80]\,M_\odot$ we renormalise both to unit integral, thereby removing the amplitude (already absorbed into $f_{\rm s}$) and isolating the morphological agreement.
All quantities are evaluated sample-by-sample over the posterior; we quote medians with $90\%$ credible intervals (Extended Data Figure~\ref{fig:ratio_function}).

\paragraph{Bhattacharyya overlap and Jensen--Shannon similarity.}
We adopt the Bhattacharyya coefficient \cite{Bhattacharyya1943}
\begin{equation}
\mathrm{BC}(p,q)=\int \sqrt{p(m)\,q(m)}\;dm \in [0,1],
\end{equation}
where $\mathrm{BC}=1$ corresponds to identical normalised spectra. Taking $\mathrm{BC}>0.9$ as the threshold for strong shape consistency, the median rises monotonically from $\mathrm{BC}=0.88^{+0.08}_{-0.20}$ in $[10,80]\,M_\odot$ to $0.95^{+0.03}_{-0.07}$ in $[20,80]\,M_\odot$, $0.97^{+0.03}_{-0.06}$ in $[30,80]\,M_\odot$, and $0.98^{+0.02}_{-0.06}$ in $[40,80]\,M_\odot$.
The criterion is therefore satisfied for all $m_{\rm w,low}\!\gtrsim\!20\,M_\odot$; the slightly lower value in $[10,80]\,M_\odot$ traces the known low-mass excess of $p(m)$ rather than a peak mismatch.
The Jensen--Shannon similarity $1-\mathrm{JSD}/\ln 2$ \cite{Lin1991} follows the same trend, increasing from $0.85^{+0.10}_{-0.22}$ in $[10,80]\,M_\odot$ to $0.93^{+0.05}_{-0.09}$, $0.95^{+0.04}_{-0.08}$, and $0.97^{+0.03}_{-0.08}$ in the successively narrower windows.
The progressive tightening of the credible intervals as $m_{\rm w, low}$ increases (Extended Data Figure~\ref{fig:ratio_function}, left and center) confirms that the spread at low $m_{\rm w,low}$ is driven almost entirely by the low-mass tail.

\paragraph{Wasserstein-1 (peak alignment).}
Because $\mathrm{BC}$ is relatively insensitive to small horizontal shifts, we complement it with the 1-Wasserstein (earth-mover) distance \cite{Villani2009},
\begin{equation}
W_1(p,q)=\int_0^1 \big|\,P^{-1}(u)-Q^{-1}(u)\,\big|\,du \quad [M_\odot],
\end{equation}
with $P,Q$ the cumulative distributions; $W_1$ is the mean mass by which $p$ must be transported onto $q$ and thus directly probes peak alignment.
Adopting $W_1\!\ll\!\Delta m$ (a small fraction of the $70\,M_\odot$ window width) as the alignment criterion, we find $W_1$ decreasing from $4.1^{+4.9}_{-2.6}\,M_\odot$ ($[10,80]\,M_\odot$) and $4.2^{+5.3}_{-2.6}\,M_\odot$ ($[20,80]\,M_\odot$) to $3.0^{+3.6}_{-1.8}\,M_\odot$ ($[30,80]\,M_\odot$) and $2.3^{+2.8}_{-1.6}\,M_\odot$ ($[40,80]\,M_\odot$).
In every window $W_1\lesssim6\%$ of the window width, confirming that the two spectra share a common peak location and that the residual difference is confined to the low-mass tail. 
Taken together, the high Bhattacharyya overlap, the high Jensen--Shannon similarity, and the small Wasserstein distance all support a hierarchical-merger origin for the high-spin massive subpopulation.

\begin{figure}
\centering  
\includegraphics[width=0.8\linewidth]{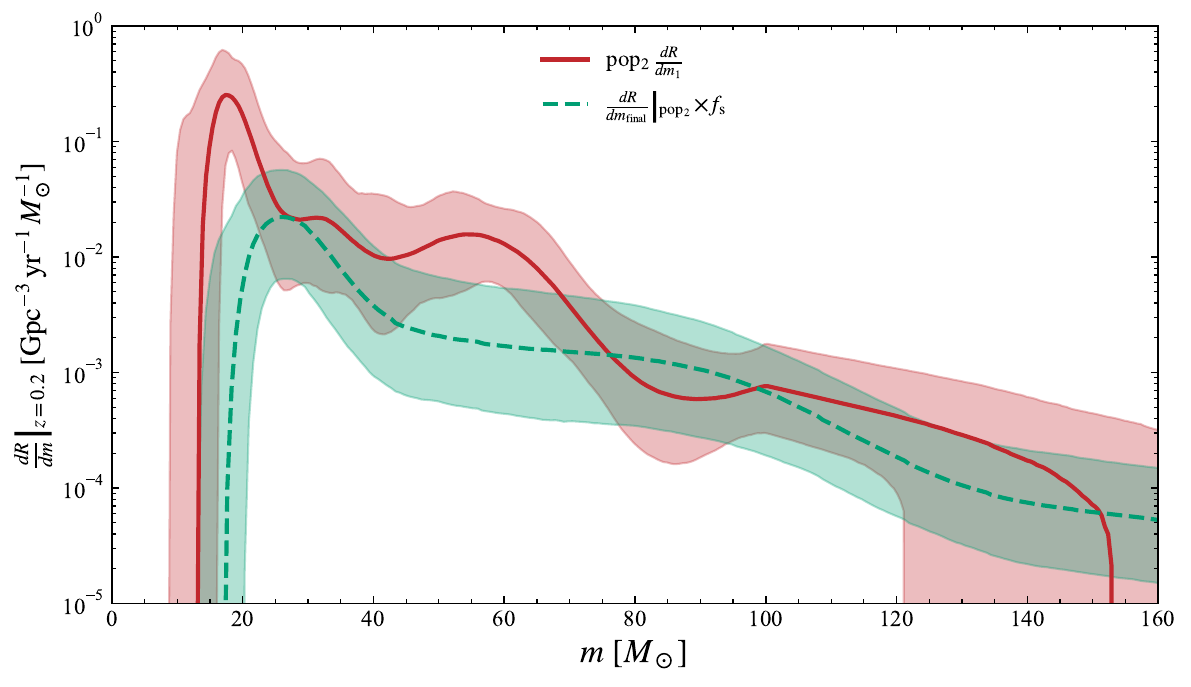}
\caption{The comparison of the reconstructed mass distribution of the second subpopulation (in red) with its remnant population scaled with a fraction of $f_{\rm s}= 0.32^{+0.25}_{-0.16}$.}
\label{fig:pop2}
\end{figure}

\begin{figure}
\centering  
\includegraphics[width=0.99\linewidth]{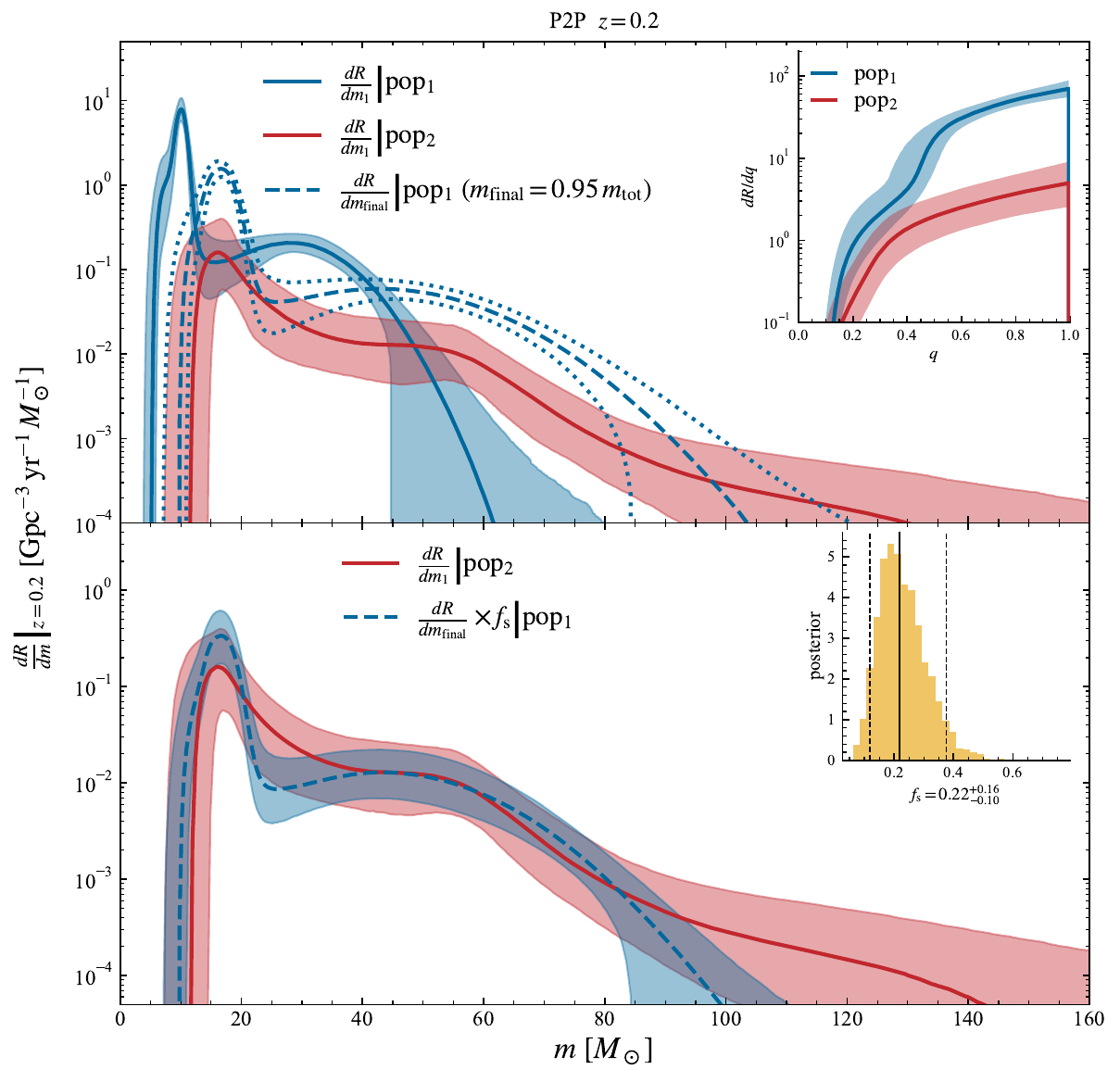}
\caption{Similar to Figure~\ref{fig:m1_mf}, but for the parametric mass function (\textsc{PowerLaw+2Peak}).}
\label{fig:P2P}
\end{figure}

\begin{figure}
\centering  
\includegraphics[width=0.99\linewidth]{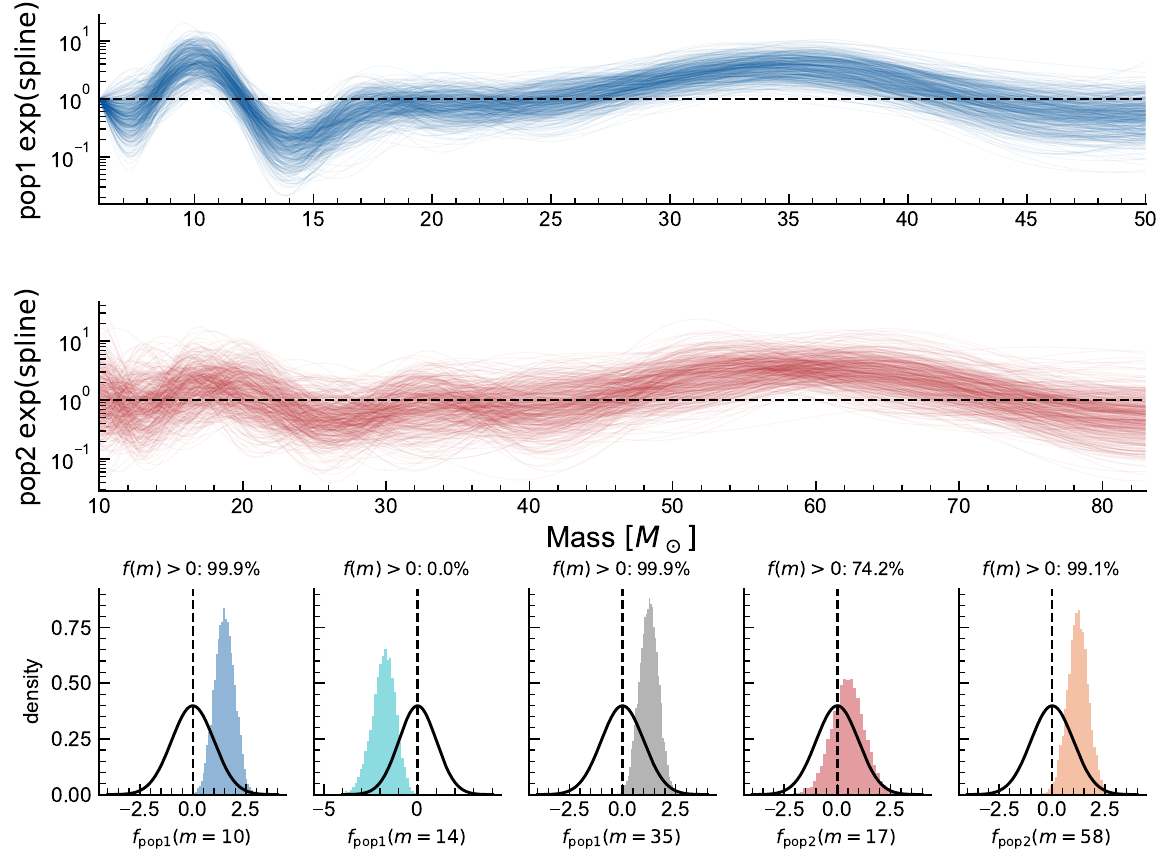}
\caption{Spline functions of the two subpopulations. The over-densities located at $\sim10\,M_\odot$, $\sim35\,M_\odot$ (first subpopulation) and $\sim58\,M_\odot$ (second subpopulation) are supported at the $>99\%$ credible level.}
\label{fig:spline}
\end{figure}

\begin{table}[t]
\centering
\caption{Summary of hyper-parameters and priors for the fiducial model} 
\label{tab:prior_xeff_polar_xp}
\begin{tabular}{lcc}
\hline\hline
Parameter & Description & Prior \\
\hline
\multicolumn{3}{c}{\bf Rate evolution} \\
$R_0$ & Local merger-rate density normalization & $U(1,100)$ \\
$\gamma$ & Redshift evolution index in $(1+z)^\gamma$ & $U(0,5)$ \\
\hline
\multicolumn{3}{c}{\bf Mass-ratio model} \\
$\beta_1$ & Power-law index of $p_1(m_2\mid m_1)\propto m_2^{\beta_1}$ (subpop.\ 1) & $U(0,8)$ \\
$\mu_q$ & Mean of $p_2(m_2\mid m_1)$ in units of $m_1$ (subpop.\ 2) & $U(0,1)$ \\
$\sigma_q$ & Std of $p_2(m_2\mid m_1)$ in units of $m_1$ (subpop.\ 2) & $U(0.01,1)$ \\
\hline
\multicolumn{3}{c}{\bf Primary-mass distribution: subpopulation 1} \\
$\alpha_1$ & Power-law slope & $U(0,8)$ \\
$m_{\min,1}\,[M_\odot]$ & Minimum mass & $U(2,10)$ \\
$m_{\max,1}\,[M_\odot]$ & Maximum mass & $U(30,100)$ \\
$\delta_1\,[M_\odot]$ & Low-mass smoothing scale & fixed to $5$ \\
$\{n_j\}_{j=1}^{15}$ & Spline-node amplitudes in $\exp[f_1(m_1)]$ & $n_1=n_{15}=0$; $n_{2\ldots14}\sim \mathcal{N}(0,1)$ \\
\hline
\multicolumn{3}{c}{\bf Spin distribution: subpopulation 1} \\
$\mu_{x,1}$ & Mean of $\chi_{\rm eff}$ truncated Gaussian & $U(-0.2,0.2)$ \\
$\lg\sigma_{x,1}$ & $\log_{10}$ std of $\chi_{\rm eff}$ & $U(-1.5,0.5)$ \\
$\mu_{y,1}$ & Baseline mean of $\chi_{\rm p}$ & $U(0,1)$ \\
$\lg\sigma_{y,1}$ & $\log_{10}$ std of $\chi_{\rm p}$ & $U(-1.5,0.5)$ \\
$\rho_1$ & Linear correlation coefficient in $\mu_{y,1}+\rho_1\chi_{\rm eff}$ & fixed to $0$ \\
\hline
\multicolumn{3}{c}{\bf Primary-mass distribution: subpopulation 2} \\
$\alpha_2$ & Power-law slope & $U(0,8)$ \\
$m_{\min,2}\,[M_\odot]$ & Minimum mass & $U(5,40)$ \\
$m_{\max,2}\,[M_\odot]$ & Maximum mass & $U(100,200)$ \\
$\delta_2\,[M_\odot]$ & Low-mass smoothing scale & fixed to $5$ \\
$\{o_j\}_{j=1}^{12}$ & Spline-node amplitudes in $\exp[f_2(m_1)]$ & $o_1=o_{12}=0$; $o_{2\ldots11}\sim \mathcal{N}(0,1)$ \\
\hline
\multicolumn{3}{c}{\bf Spin distribution: subpopulation 2 (polar semi-annulus)} \\
$r_2$ & Mixture fraction of subpopulation 2 & $U(0,1)$ \\
$l_{\chi_{\rm eff}}$ & Ellipse scale along $\chi_{\rm eff}$ (via $x=\chi_{\rm eff}/l_{\chi_{\rm eff}}$) & $U(0,1)$ \\
$l_{\chi_{\rm p}}$ & Ellipse scale along $\chi_{\rm p}$ (via $y=\chi_{\rm p}/l_{\chi_{\rm p}}$) & $U(0,1)$ \\
$\lg\sigma_{\chi}$ & $\log_{10}$ radial width parameter $\sigma_r$ & $U(-1.5,0.5)$ \\
$t_{\min}$ & Lower angular bound & $U(-\pi/2,\pi/2)$ \\
$t_{\max}$ & Upper angular bound & $U(-\pi/2,\pi/2)$ \\
$t_{\max}-t_{\min}$ & Constraint & $\in(0,\pi)$ \\
$\{n_{x,j}\}_{j=1}^{6}$ & Angular-spline amplitudes in $\exp[f(t)]$ & $n_{x,j}\sim \mathcal{N}(0,1)$ \\
\hline\hline
\end{tabular}
\end{table}

\begin{table}[htpb]
\centering
\caption{Model comparison}\label{tab:BF}
\begin{tabular}{lccc}
\hline
\hline
Model     &  $\ln{\mathcal{B}}$  &  ${\mathcal{B}}$ \\
\hline
P2P(pop$_1$)+SPL(pop$_2$)  & 0 & 1  \\
P2P(pop$_1$)+P2P(pop$_2$)  & 3  & 20  \\
Fiducial (Semi-parametric) model & 6 & 403\\
Single-population model & -35 & $e^{-35}$\\
\hline
\end{tabular}
\\
\begin{tabular}{l}
Note: All the (ln) Bayes factors are relative to the model with\\
P2P + SPL for the mass function of two subpopulations.
\end{tabular}
\end{table} 

\begin{addendum}
\item[Acknowledgements] We thank Shi-Jie Gao, Wen-Yu Xin, and Ming-Zhe Han for helpful discussion. This work is supported by the National Natural Science Foundation of China (No. 12588101, No. 12233011, No. 12503059, No. 12203101, No. 12303056), the New Cornerstone Science Foundation through the XPLORER PRIZE, and the Priority Research Program of the Chinese Academy of Sciences (No. XDB0550400).
This research has made use of data and software obtained from the Gravitational Wave Open Science Center (https://www.gw-openscience.org), a service of LIGO Laboratory, the LIGO Scientific Collaboration and the Virgo Collaboration. LIGO is funded by the U.S. National Science Foundation. Virgo is funded by the French Centre National de Recherche Scientifique (CNRS), the Italian Istituto Nazionale della Fisica Nucleare (INFN) and the Dutch Nikhef, with contributions by Polish and Hungarian institutes.

\item[Competing Interests] The authors declare no competing interests.

\item[Author Information] Correspondence and requests for materials should be addressed to Y.Z.F.~(yzfan@pmo.ac.cn).
\end{addendum}

\newpage
\renewcommand{\refname}{References}
\bibliographystyle{sn-nature}
\bibliography{refs.bib}

%%%%%%%%%%%%%%%%%%%%%%%%%%%%%%%%%%%%%%%%%%%%%%%%%%%%%%%%%%%%%%%%%%%
%% Supplementary Information
%%%%%%%%%%%%%%%%%%%%%%%%%%%%%%%%%%%%%%%%%%%%%%%%%%%%%%%%%%%%%%%%%%%
\clearpage
\setcounter{figure}{0}
\setcounter{table}{0}
\setcounter{equation}{0}
\renewcommand{\theequation}{S\arabic{equation}}
\renewcommand{\figurename}{Supplementary Figure}
\renewcommand{\tablename}{Supplementary Table}

\begin{center}
{\bf Supplementary Information}
\end{center}

\textbf{\emph{Alternative population models}}

\paragraph{Cross-check model (spin magnitude--tilt).}
As an independent cross-check we analyze the population in the $(m_1,m_2,a_1,a_2,\cos\theta_1,\cos\theta_2,z)$ space, again with a two-component mixture \cite{2025ApJ...987...65L,2026SCPMA..6999562W}.
The two compact objects in each binary are assumed to be drawn from a common single-object distribution, paired through a power-law factor in the mass ratio,
\begin{equation}
\begin{split}
p(m_1,m_2,&a_1,a_2,\cos\theta_1,\cos\theta_2,z\mid\Lambda)\\
\propto{}&~ p(z\mid\gamma)\,
\left(\frac{m_2}{m_1}\right)^{\beta}
\mathcal{P}(m_1,a_1,\cos\theta_1)\\
&\times \mathcal{P}(m_2,a_2,\cos\theta_2),
\qquad m_2<m_1,
\end{split}
\end{equation}
where $p(z\mid\gamma)\propto(1+z)^\gamma$ and the joint normalization over $(m_1,m_2)$ is computed numerically.
The single-object density is itself a two-component mixture,
\begin{equation}
\begin{split}
\mathcal{P}(m,a,\cos\theta)
={}&(1-r_2)\,p_1(m)\,p_1(a)\,p_1(\cos\theta)\\
&+ r_2\,p_2(m)\,p_2(a)\,p_2(\cos\theta).
\end{split}
\end{equation}
For each subpopulation $i\in\{1,2\}$ the primary-mass term $p_i(m)$ has the same spline-modulated power-law form as in our primary analysis (with 12 knots), the spin magnitude follows a truncated Gaussian
\begin{equation}
p_i(a)=\mathcal{N}(a\mid \mu_{a,i},\sigma_{a,i})_{[a_{\min,i},a_{\max,i}]},
\end{equation}
and the spin tilt follows a mixture of an aligned truncated Gaussian and an isotropic component,
\begin{equation}
\begin{split}
p_i(\cos\theta)={}&\zeta_i\,
\mathcal{N}(\cos\theta\mid \mu_{t,i},\sigma_{t,i})_{[-1,1]}\\
&+(1-\zeta_i)\,\mathcal{U}(-1,1).
\end{split}
\end{equation}
To keep the two subpopulations separated in spin magnitude we impose $a_{\min,2}\simeq a_{\max,1}$ as a soft constraint. The priors of this cross-check model are summarized in Supplementary Table~\ref{tab:prior_double_mact}. The results are shown in Supplementary Figure~\ref{fig:mact}.

\paragraph{Total-mass cross-check model.}
As a further cross-check, we analyze the population directly in the $(M_{\rm tot},q,z)$ space, where $M_{\rm tot}=m_1+m_2$ is the source-frame total mass and $q=m_2/m_1$ is the mass ratio.
Here we adopt a single-component model and do not model the spin distribution, assuming the spin components to be uniformly distributed.
The population reads
\begin{equation}
p(M_{\rm tot},q,z\mid\Lambda)
= p(z\mid\gamma)\,p(M_{\rm tot})\,p(q),
\end{equation}
and the corresponding density in $(m_1,m_2)$ is obtained through the Jacobian $|\partial(M_{\rm tot},q)/\partial(m_1,m_2)|$.

The total-mass distribution has the same spline-modulated power-law form as in our primary analysis,
\begin{equation}
\begin{split}
p(M_{\rm tot})\propto{}& M_{\rm tot}^{-\alpha}\,
S(M_{\rm tot}\mid m_{\min},\delta)\,
\exp\!\big[f(M_{\rm tot})\big],\\
& M_{\rm tot}\in(m_{\min},m_{\max}),
\end{split}
\end{equation}
where $S$ is the smooth turn-on defined in the Methods and $f(M_{\rm tot})$ is a natural cubic spline specified by amplitudes $\{n_j\}_{j=1}^{15}$ at $15$ knots logarithmically spaced in $[10,200]\,M_\odot$ (with $n_1=n_{15}=0$).
The normalization is computed numerically.
The mass ratio follows a simple power law,
\begin{equation}
p(q)\propto q^{\beta},\qquad q\in[0.1,1].
\end{equation}
In this model, we do not consider the spin distribution. The inferred total-mass distribution is converted to the final-mass distribution with the approximate function $m_{\rm final}=0.95M_{\rm tot}$, and then compared to the primary-mass distribution of the high-spin subpopulation inferred with the fiducial model, see Supplementary Figure~\ref{fig:mfinal_direct}.

\begin{table}[t]
\centering
\caption{Hyper-parameters and priors for the cross-check model.}
\label{tab:prior_double_mact}
\begin{tabular}{lcc}
\hline\hline
Parameter & Description & Prior \\
\hline
\multicolumn{3}{c}{\bf Rate evolution and pairing} \\
$R_0$ & Local merger-rate density normalization & $U(0,100)$ \\
$\gamma$ & Redshift evolution index in $(1+z)^\gamma$ & $U(-2,7)$ \\
$\beta$ & Pairing power-law index in $(m_2/m_1)^\beta$ & $U(-8,8)$ \\
\hline
\multicolumn{3}{c}{\bf Mass distribution} \\
$\alpha_i$ & Power-law slope ($i=1,2$) & $U(-8,8)$ \\
$m_{\min,1},\,m_{\min,2}\,[M_\odot]$ & Minimum mass & $U(2,10),\ U(2,40)$ \\
$m_{\max,1},\,m_{\max,2}\,[M_\odot]$ & Maximum mass & $U(20,100),\ U(100,200)$ \\
$\delta_{m,i}\,[M_\odot]$ & Low-mass smoothing scale & $U(0,10)$ \\
$\{n_j\}_{j=2}^{11},\{o_j\}_{j=2}^{11}$ & Spline-node amplitudes & $\sim\mathcal{N}(0,1)$ \\
$\{n_j\}_{j=1,12},\{o_j\}_{j=1,12}$ & First and last Spline-node amplitudes & 0 \\
$r_2$ & Mixture fraction of subpopulation 2 & $U(0,1)$ \\
\hline
\multicolumn{3}{c}{\bf Spin magnitude} \\
$\mu_{a,1},\,\mu_{a,2}$ & Mean of $a$ truncated Gaussian & $U(0,1)$ \\
$\sigma_{a,1},\,\sigma_{a,2}$ & Std of $a$ & $U(0.05,0.5)$ \\
$a_{\min,1},\,a_{\max,1}$ & Bounds, subpop.\ 1 & $0,\ U(0.25,0.75)$ \\
$a_{\min,2},\,a_{\max,2}$ & Bounds, subpop.\ 2 & $U(0.25,0.75),\ 1$ \\
$a_{\min,2}-a_{\max,1}$ & Separation constraint & $>0$ \\
\hline
\multicolumn{3}{c}{\bf Spin tilt} \\
$\mu_{t,1},\,\mu_{t,2}$ & Mean of $\cos\theta$ Gaussian & $U(-1,1)$ \\
$\sigma_{t,1},\,\sigma_{t,2}$ & Std of $\cos\theta$ & $U(0.1,4)$ \\
$\zeta_1,\,\zeta_2$ & Aligned (vs.\ isotropic) fraction & $U(0,1)$ \\
\hline\hline
\end{tabular}
\end{table}

\begin{figure}
\centering  
\includegraphics[width=0.99\linewidth]{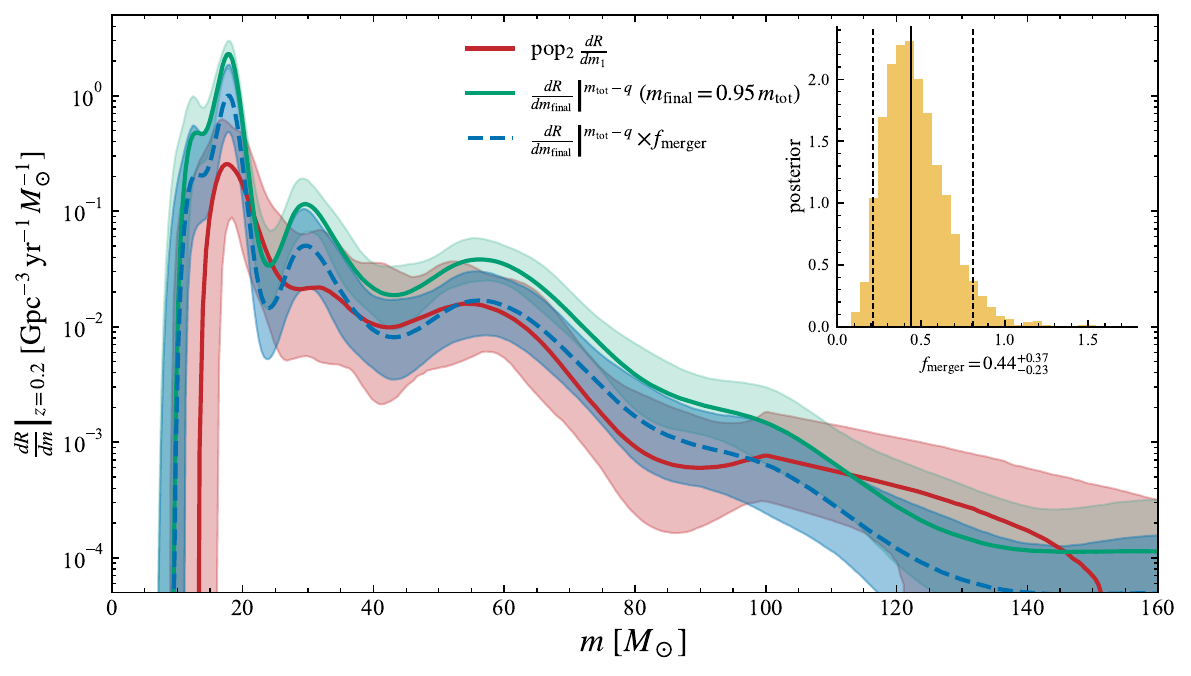}
\caption{Primary-mass distribution of the second subpopulation compared to the final-mass distribution of all the population inferred with the total-mass cross-check model. Solid lines: medians; shaded regions: 90\% credible intervals.
The features of the two subpopulations are still consistent with each other.}
\label{fig:mfinal_direct}
\end{figure}

\begin{figure}
\centering  
\includegraphics[width=0.99\linewidth]{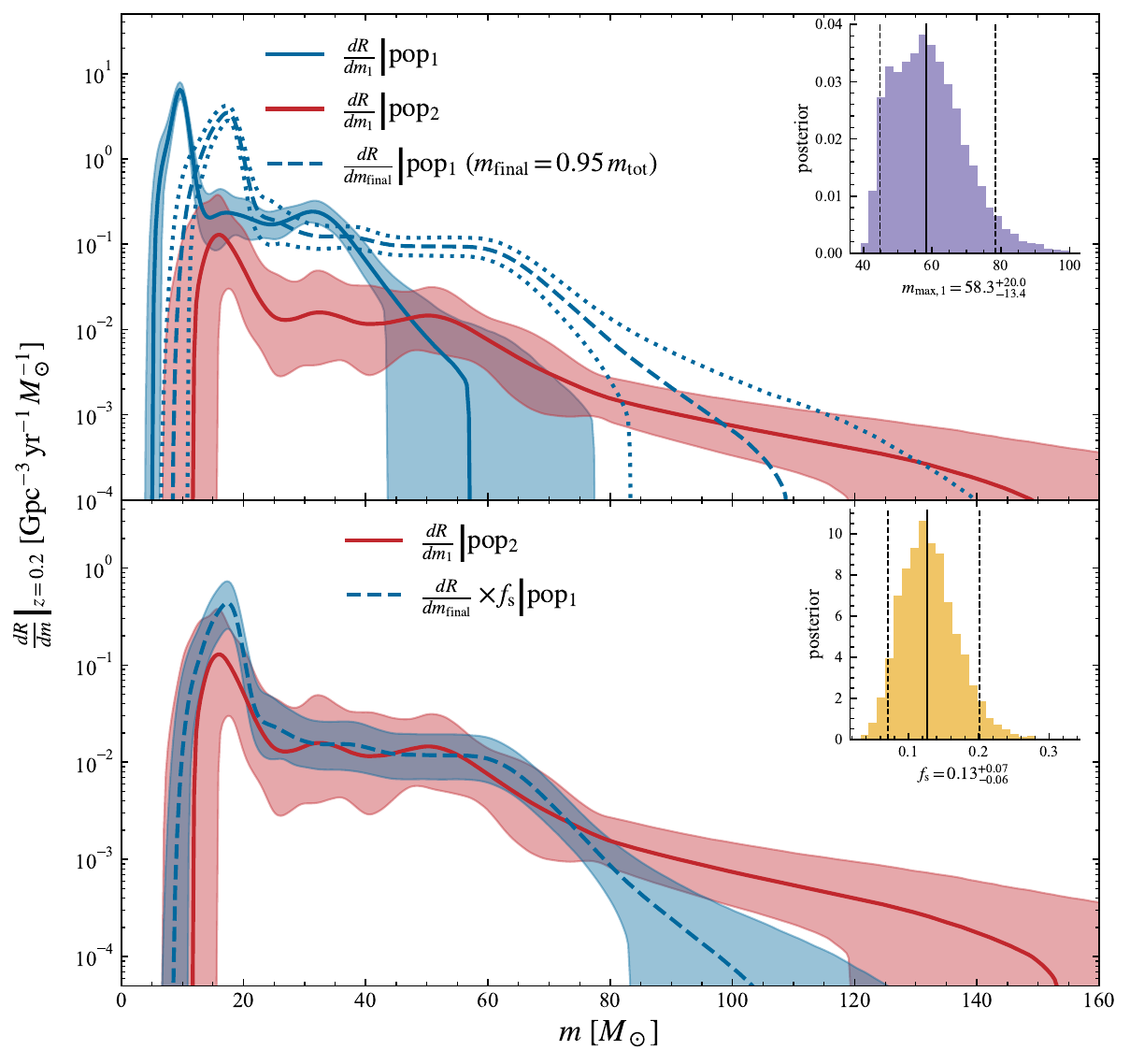}
\caption{Similar to Figure~\ref{fig:m1_mf}, but for the alternative $(m_1,m_2,a_1,a_2,\cos\theta_1,\cos\theta_2,z)$ model.}
\label{fig:mact}
\end{figure}

\textbf{\emph{The properties of subpopulations}}

\begin{figure}
\centering  
\includegraphics[width=0.8\linewidth]{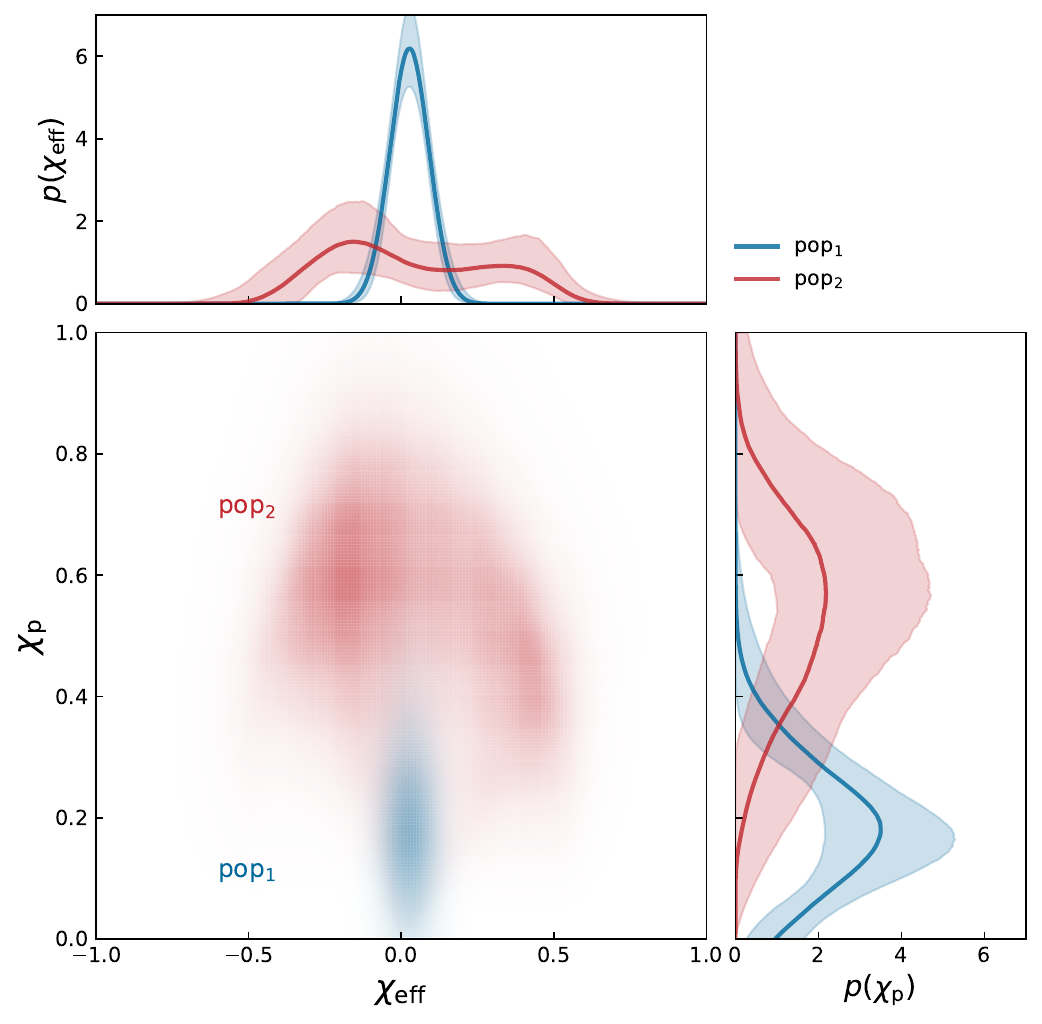}
\caption{The reconstructed spin ($\chi_{\rm eff}$ vs. $\chi_{\rm p}$) distributions of the two subpopulations.}
\label{fig:Spin_dist}
\end{figure}

\begin{figure}
\centering  
\includegraphics[width=0.8\linewidth]{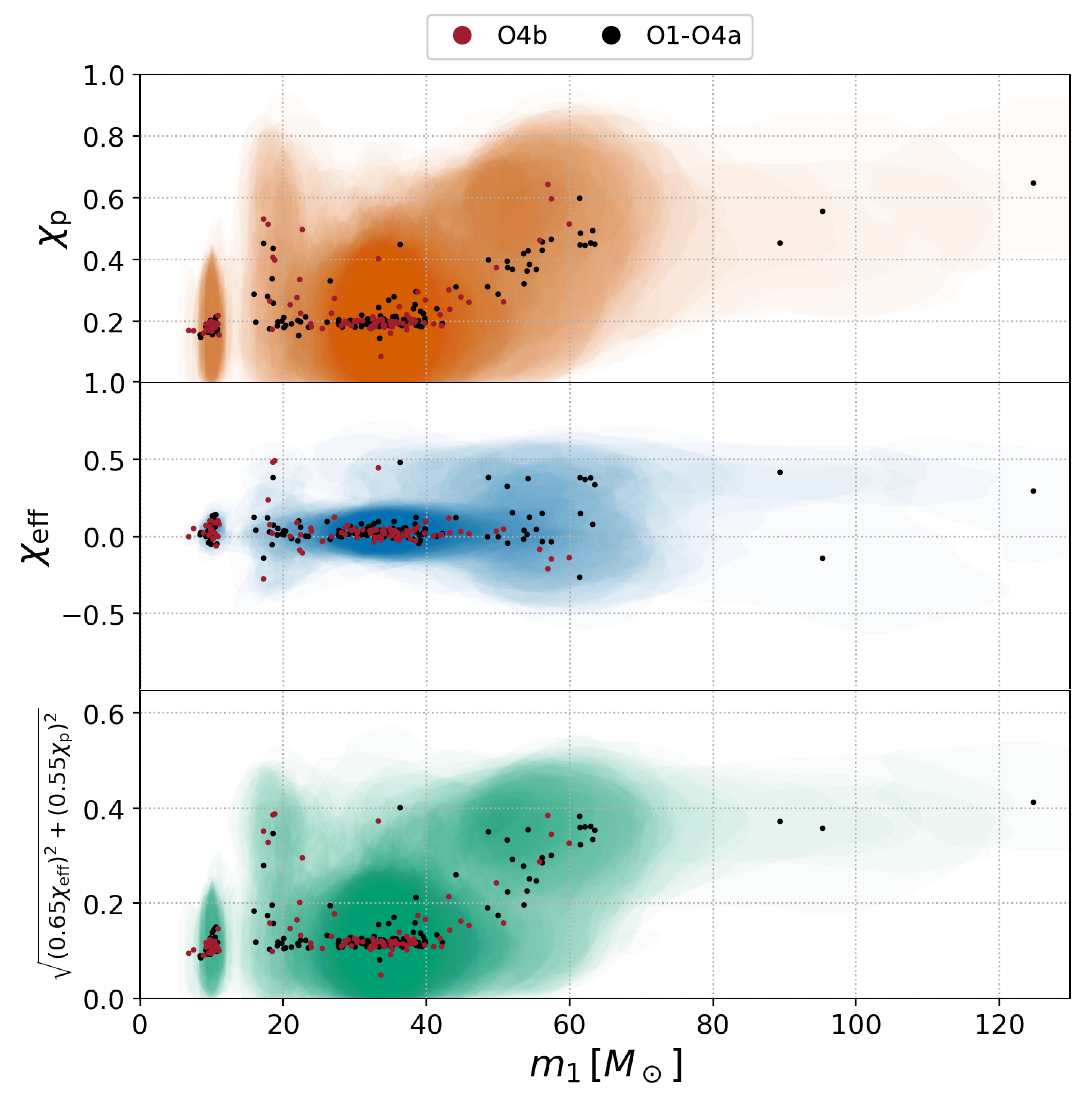}
\caption{Population-informed posterior of primary mass, effective spin, effective precession, and combined spin for GWTC-5 events.
Note the stars are for the median informed samples showing the approximate distribution tendency. The darkness of the contours indicates the density distribution.
}
\label{fig:informed}
\end{figure}

\begin{figure}
\centering  
\includegraphics[width=0.5\linewidth]{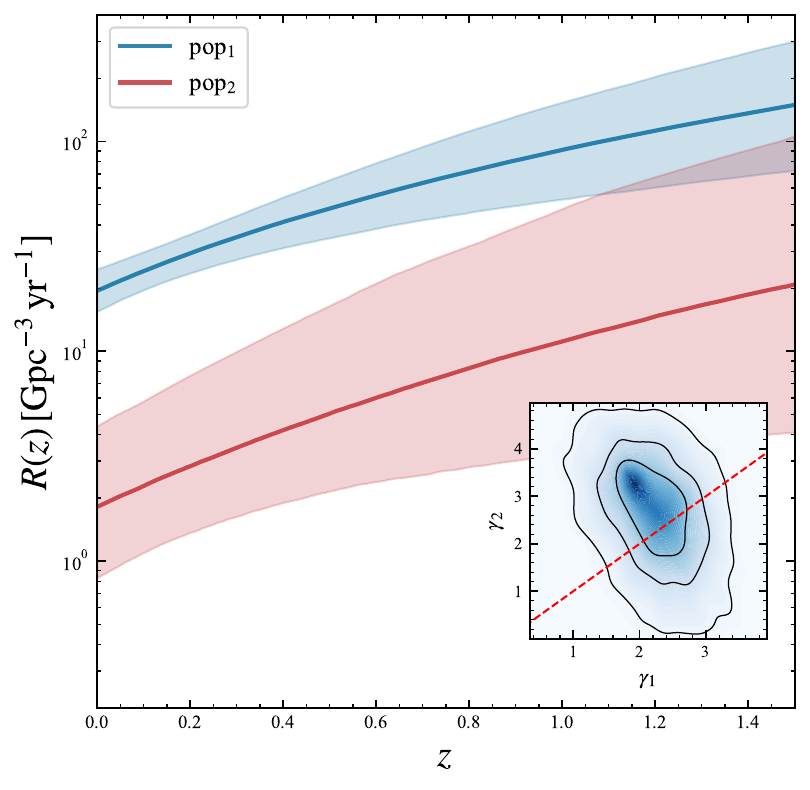}
\caption{The redshift distribution of the BBH subpopulations, and the corresponding spectral index.}
\label{fig:gamma}
\end{figure}

Supplementary Figure~\ref{fig:Spin_dist} shows the separation of the two subpopulation in the $\chi_{\rm eff}-\chi_{\rm p}$ 2-D space.
For the high-spin subpopulation, a prominent semi-elliptical shape appears with two semi-axes of $l_{\chi_{\rm eff}}=0.53^{+0.19}_{-0.2}$ and $l_{\chi_{\rm p}} = 0.62^{+0.18}_{-0.24}$.
We therefore construct another indicator $\chi_{\rm sq}\equiv \sqrt{(0.65\chi_{\rm eff})^2+(0.55\chi_{\rm p})^2}$ to observe the separation of the two black hole populations within the mass range.
It shows that the two subpopulations are more clearly separated in the $m_1-\chi_{\rm sq}$ plane than in the $m_1-\chi_{\rm p}$ or $m_1-\chi_{\rm eff}$ plane, see Supplementary Figure~\ref{fig:informed} (note that the contours, but not the points/stars are for visual inspection).
Such a separation is similar to those in the component-mass vs spin-magnitude plane \cite{2025arXiv250923897L, 2026SCPMA..6999562W}.
This separation in the spin plane is what makes the two mass functions individually identifiable.

Refs.~\cite{2024ApJ...975...54G, 2026ApJ..1001L..40F} find that hierarchical mergers evolve faster than the first-generation mergers with GWTC-3 and GWTC-4, respectively. 
However, we find no evidence that hierarchical mergers evolve faster using an extend model which incorporates independent rate-evolution slopes for the two subpopulations ($\gamma_1$ and $\gamma_2$).
We find the hierarchical-merger rate evolves similar to that of the first-generation mergers with redshift, see Supplementary Figure~\ref{fig:gamma}.

\textbf{\emph{The impact of the variance threshold for likelihood estimation}}

\begin{figure}
\centering
\includegraphics[width=0.99\linewidth]{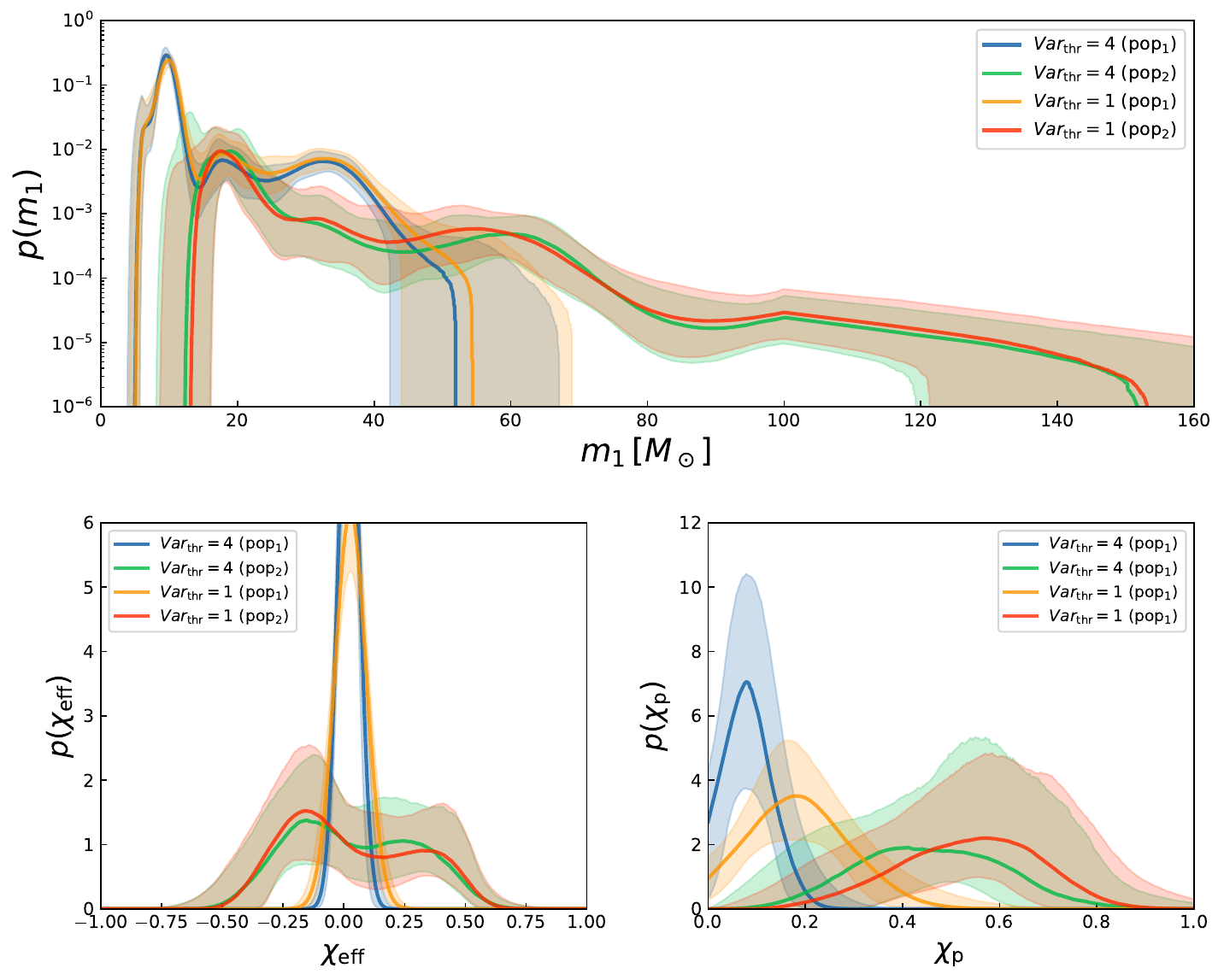}
\caption{Comparing the mass and spin distributions of subpopulations, inferred with different thresholds for likelihood uncertainty of Monte Carlo integration with finite posterior samples and injection samples.}
\label{fig:varth}
\end{figure}

Monte Carlo uncertainty in the likelihood evaluation may influence the hierarchical inference \cite{2023MNRAS.526.3495T}.
To check whether such uncertainty has impact on our results, we repeat our analysis with a more permissive cutoff $Var_{\rm thr}=4$, and compare to the results in the main text, i.e., inferred with $Var_{\rm thr}=1$.
It turns out that the mass distributions are well consistent between the two results (see Supplementary Figure~\ref{fig:varth}), demonstrating that the main finding of this work is reliable.

\end{document}